\documentclass[12pt,thmsa]{article}\usepackage{amsmath}

\usepackage{amssymb}
\usepackage{sw20lart}

\input{tcilatex}

\begin{document}

\title{Spin Current and Shot Noise in Single-Molecule Quantum Dots with a
Phonon Mode }
\author{Hui Yu and J.-Q. Liang \\
Institute of Theoretical Physics, Shanxi University, Taiyuan, Shanxi 030006,%
\\
China}
\maketitle

\begin{abstract}
In this paper we investigate the spin-current and its shot-noise spectrum in
a single-molecule quantum dot coupled with a local phonon mode. We pay
special attention on the effect of phonon on the quantum transport property.
The spin-polarization dependent current is generated by a rotating magnetic
filed applied in the quantum dot. Our results show the remarkable influence
of phonon mode on the zero-frequency shot noise. The electron-phonon
interaction leads to sideband peaks which are located exactly on the integer
number of the phonon frequency and moreover the peak-height is sensitive to
the electron-phonon coupling.
\end{abstract}

PACS : 73.23.-b; 85.75.-d ;73.50.Td ;85.65.+h

\section{INTRODUCTION}

Modern nanotechnology has provided a possibility to fabricate electronic
devices in which the acting element is a single, organic molecule. This
leads to a growing interest in the study of transport properties of
molecular devices\cite{1,2,3}. Such a device may be modeled as a `quantum
dot' (QD) weakly coupled to the macroscopic charge reservoirs. In addition
to their practical applications these artificial ,tunable devices such as
electronic components\cite{1}, coulomb blockade structures\cite{4}, diodes%
\cite{5} or switching devices with high negative differential resistance\cite%
{6} , are important for understanding the basic physics including the
many-body and the size effects. In contrast to the semiconductor QDs,
molecules ( the linear size is at least one order smaller than that of the
former) are intrinsically different from semiconductor nanostructures. The
devices with molecules may lead to new physics especially when electrons are
added or removed from a single-molecule. Since the molecular material
possesses much smaller elastic parameters it is very easy to excite their
internal, vibrational degrees of freedom (phonon modes) \cite{6,7,8,9,10,11}
when electrons are incident upon the molecules through a tunnel junction.
Thus molecules react inevitably back to the tunnel electrons even at low
temperature. This phenomenon has now provoked a large amount of experimental 
\cite{3,8,10,11,12,13,14,15}investigations in the problem of transport
through mesoscopic systems with electron- phonon coupling. Inelastic
scattering effects have been observed directly in measurements of the
differential conductance of molecules adsorbed on metallic substrates with
scanning tunneling microscopy (STM)\cite{3}. In a series of pioneering
experiments by Park \textit{et al}\cite{8}\ \textit{\ } it was shown that
the current through a single $C_{60}$ molecule was strongly influenced by
the vibrational mode. Zhitenev \textit{et al}.\cite{11} have also
demonstrated that the low-bias conductance of molecules is dominated by
resonant tunneling through coupled electronic and vibrational levels. There
has been a number of theoretical efforts \cite{14,15,16,17,18,19,20,21,22,24}
focused on the effects of electron-phonon(E-PH) coupling in mesoscopic
systems with basic models capturing the essential physics and standard
methods. Various aspects of the electron-phonon/vibron interaction effect on
the tunneling through molecule QDs have been studied by many authors.

On the other hand, motivated by the easy control of electron-spin as well as
the remarkably long coherence time, the spin-polarization dependent
transports in open QDs \cite{26,27,28} have attracted considerable
attentions. These spin-source devices not only exhibit the new fundamental
physics but also are of promising applications in the emerging technologies
of spintronics and quantum information\cite{29}. A pure spin current has
been reported by direct optical injection without generating a net charge
current\cite{30}. Theoretically, there are number of mechanisms proposed to
produce pure spin-current\cite{31,32,33} using techniques of ferromagnetic
resonance in a ferromagnetic-normal-metal\cite{31} or electron spin
resonance(ESR) in a QD-based system with sizable Zeeman splitting\cite{32}.

Moreover, a modern trend of the transport studies in mesocopic systems is
toward not only to considering the transport characteristics of a given
device but also to examining the noise properties. Due to the discrete
nature of charge carriers, electrical current through a conductor is subject
to time-dependent fluctuation around its mean value which manifests the
consequence of the quantization of the charge carriers and is usually
referred to as the shot noise in literature\cite{41}. Shot noise defined as
the mean-square fluctuations of the current flowing through a given terminal
at zero temperature is of great importance and interest because the spectrum
of shot noise contains additional information about the interactions which
the conduction electrons undergo\cite{40} beyond the mean-current properties
and can be used to discern different mechanisms resulting in the same mean
current. For example, shot noise experiments can determine the kinetics of
electron and reveal information of the correlation of electronic wave
functions. Thus it has been extensively studied in a wide variety of systems%
\cite{40}. Furthermore, the spin-resolving current correlation is more
useful to describe electron correlation, because the electronic wavepacket
with opposite spins is uninfluenced by the Pauli exclusion principle and
only reflects unambiguous information about the interaction. For a two-lead
device, correlations can be formulated by quantities measured at the same
lead i.e. the auto-correlation,or by quantities measured at the two
different leads, namely, the cross-correlation. B$\ddot{u}$ttiker\cite{41}
pointed out that while cross-correlations of the charge noise can either be
positive or negative for bosons, they are necessarily negative definite for
fermions in both the equilibrium and the transport regimes. It has been well
known\cite{42} that anti-bunching in a Fermionic system gives rise to
negative definite cross-correlation for charge-current. And the conservation
of charge forces the cross correlation and auto correlation of
charge-current noise spectra to differ by just a minus sign in a two-probe
system. However for a pure spin-current the situation is very different
because of a lack of spin-current conservation due to the spin flipping. The
auto-correlation of the spin-current is surely positive definite, while the
cross-correlation is either positive or negative depending on a number of
parameters\cite{34}. The zero-frequency shot noise $S(0)$ for a classical
conductor\cite{35} is characterized by the Poisson value $%
S_p(0)=2e\left\langle I\right\rangle $ where $\left\langle I\right\rangle $
is the average current, while the shot noise in a non-interacting mesoscopic
conductor is always reduced by the Pauli exclusion in comparison with the
Poisson value. Of course, shot noise is also influenced by other factors
such as electron-electron and electron-phonon interactions.

Motivated by the achievements in the single-molecule and spin-current
experiments we in this paper will, based on the ESR mechanism \cite{34,36}of
generating net spin-current\cite{25}, investigate the electron-phonon
effects on the spin-current and its shot noise which have not yet been
studied theoretically for molecular dots. We use the Keldysh nonequilibrium
Green function technique to calculate the spin current and shot noise
through a single-molecule coupled to electron reservoirs for the first time
and focus on the effect of inelastic scattering process. The ESR-type model
with a single dispersionless phonon mode is employed to address the
vibrational degrees of freedom in the molecular dot. All other complexity of
real molecular devices, apart from interaction with a single longitudinal
optical(LO) phonon localized on the molecule, is ignored.

\section{MODEL AND HAMILTONIAN}

The model system under consideration is illustrated in Fig. 1, which
consists of a molecule of one relevant level coupled with a single
(Einstein) phonon mode and two leads which we label as `left' and `right'. A
time-dependent (rotating) magnetic field $\mathbf{B}(t)$ is applied in the
molecule dot to flip the spin of electron. Also a gate electrode is
capacitively attached to the dot to tune the energy level. The total
Hamiltonian of the system is written as 
\begin{equation}
H=H_L+H_R+H_P+H_D+H^{\prime }(t)+H_T,  \label{1}
\end{equation}
where

\begin{equation}
H_L+H_R=\underset{k\sigma ,\alpha =L,R}{\sum }\varepsilon _kC_{k\alpha
\sigma }^{+}C_{k\alpha \sigma }  \label{2}
\end{equation}
\begin{equation}
H_P=w_0a^{+}a  \label{3}
\end{equation}
\begin{equation}
H_D=\underset{\sigma }{\sum }[(\varepsilon +\sigma B_0\cos \theta )+\lambda
(a+a^{+})]d_\sigma ^{+}d_\sigma  \label{4}
\end{equation}
\begin{equation}
H^{\prime }(t)=r[\exp (-iwt)d_{\uparrow }^{+}d_{\downarrow }+\exp
(iwt)d_{\downarrow }^{+}d_{\uparrow }]  \label{5}
\end{equation}
with $r=B_0\sin \theta $%
\begin{equation}
H_T=\underset{k,\sigma ,\alpha =L,R}{\sum }[T_{k\alpha }C_{k\alpha \sigma
}^{+}d_\sigma +c.c].  \label{6}
\end{equation}
The first two terms $H_L$ and $H_R$ of eq.(1) are respectively the
Hamiltonians for electrons in the left and right non-interacting metallic
leads, where $C_{k\alpha \sigma }^{+}(C_{k\alpha \sigma })$ are the creation
(annihilation) operators of electrons with momentum $k$, spin-$\sigma $ and
energy $\varepsilon _k$ in the lead $\alpha $. Here we have set the same
chemical potential for both leads. The third term $H_X$ describes the
nondispersive, longitudinal optical (LO) phonon, where $w_0$ is the
frequency of the single phonon mode, $a^{+}(a)$ is phonon creation
(annihilation) operator. $H_D$ and $H^{\prime }(t)$ correspond to the
interaction Hamiltonias between electron and phonon in the QD which is
subjected to a time-dependent magnetic field with uniform strength, 
\[
\mathbf{B}(t)=B_0(\sin \theta \cos \omega t\text{, }\sin \theta \sin \omega t%
\text{, }\cos \theta \text{ }) 
\]
where $B_0$ is the constant field strength. Here $d_\sigma ^{+}$ $(d_\sigma
) $ are the electron-creation (annihilation) operators in the QD and $%
\varepsilon =\varepsilon _0+eV_g$ is the single energy level of the molecule
which can be controlled by the gate voltage $V_g$, where $e$\ denotes the
absolute value of electron charge.\ $\lambda $ is the coupling constant
between the electron in the molecule dot and the LO phonon mode with energy $%
w_0$ . $H_T$ represents the coupling of the molecule with leads, where the
tunneling matrix elements $T_{k\alpha }$ transfer electrons through an
insulating barrier out of the dot.

\subsection{SPIN-CURRENT AND SHOT NOISE FORMULA}

We define the spin-dependent particle current operator in the lead-$\alpha $
as ($\hbar =1$) 
\begin{equation}
\overset{\wedge }{J}_{\alpha ,\sigma \sigma ^{\prime }}\equiv \underset{k}{%
\sum }\frac{d[C_{k\alpha \sigma }^{+}C_{k\alpha \sigma ^{\prime }}]}{dt}=-i%
\underset{k}{\sum }[T_{k\alpha }C_{k\alpha \sigma }^{+}d_{\sigma ^{\prime
}}-T_{k\alpha }^{*}d_\sigma ^{+}C_{k\alpha \sigma ^{\prime }}],  \label{7}
\end{equation}
then spin-current operator of spin component $\sigma $\ is

\begin{equation}
J_{s\alpha }=\frac 12\sum_{\sigma \sigma ^{^{\prime }}}J_{\alpha ,\sigma
\sigma ^{^{\prime }}}\mathbf{\sigma }_{\sigma \sigma ^{^{\prime }}}^z
\label{8}
\end{equation}
where\textbf{\ }$\mathbf{\sigma }^z$ is Pauli matrix.

The spin-dependent current can be computed from current operator Eq. (7)

\begin{equation}
I_{\alpha ,\sigma \sigma ^{^{\prime }}}(t)\equiv \langle \overset{\wedge }{J}%
_{\alpha ,\sigma \sigma ^{^{\prime }}}\left( t\right) \rangle
=-\sum_k[T_{k\alpha }G_{d\sigma ,k\alpha \sigma ^{^{\prime }}}^{\prec
}(t,t)-T_{k\alpha }^{*}G_{k\alpha \sigma ^{^{\prime }},d\sigma }^{\prec
}(t,t)]  \label{9}
\end{equation}
where the nonequilibrium Green's functions (NEGFs) are defined as 
\[
G_{d\sigma ,k\alpha \sigma ^{^{\prime }}}^{\prec }(t,t^{^{\prime }})\equiv
i\langle C_{k\alpha \sigma ^{\prime }}^{+}(t^{^{\prime }})d_\sigma
(t)\rangle , 
\]

\[
G_{k\alpha \sigma ,d\sigma ^{^{\prime }}}^{\prec }(t,t^{^{\prime }})\equiv
i\langle d_{\sigma ^{\prime }}^{+}(t^{^{\prime }})C_{k\alpha \sigma
}(t)\rangle . 
\]
Using the Keldysh nonequilibrium Green function formalism\cite{37}we obtain
spin-dependent current

\begin{equation}
I_{\alpha ,\sigma \sigma ^{^{\prime }}}=\frac 1{2N_\tau }\int \frac{dE_1}{%
2\pi }\int \frac{dE_2}{2\pi }\Gamma _\alpha \Gamma \underset{\sigma _2}{\sum 
}G_{\sigma \sigma _2}^r(E_1,E_2)G_{\sigma _2\sigma ^{\prime
}}^a(E_2,E_1)[f(E_2)-f(E_1)]  \label{10}
\end{equation}
where 
\[
\Gamma =\underset{\alpha =L,R}{\sum }\Gamma _\alpha 
\]
is the total tunnel coupling constant which is a function of energy $E$ and $%
G_{\sigma \sigma ^{\prime }}^{r(a)}(E_1,E_2)$ are the Fourier transform of
the dot-electron retarded (advanced) Green's function 
\[
G_{\sigma \sigma ^{^{\prime }}}^{r(a)}(t,t^{^{\prime }})=\mp i\theta (\pm
t\mp t^{\prime })\left\langle \left\{ d_\sigma \left( t\right) ,d_{\sigma
^{\prime }}^{+}(t^{\prime })\right\} \right\rangle 
\]
in the presence of both the electron-phonon interaction and the tunneling
coupling between dot and leads, $\mathbf{\Gamma }_\alpha $\textbf{(}$E$%
\textbf{), }the elastic coupling to the $\alpha -$lead (referred to as the
line-width function),\textbf{\ }depends on the hopping strength and the
density of states $\rho _\alpha (E)$ in the lead-$\alpha $ according to

\begin{equation}
\Gamma _\alpha (E)\equiv 2\pi \rho _\alpha (E)\left| T_{k\alpha }(E)\right|
^2.  \label{11}
\end{equation}
In the wide-band limit\cite{25} in which the bandwidth in the leads is much
larger than both the resonance width and phonon energies, the contact
densities of states are constant in the region of the resonance. If the
hopping matrix elements also vary slowly with energy, the couplings with the
contacts $\Gamma _\alpha $ are independent of energy either. The Fermi
distribution of the lead-$\alpha $ is

\begin{equation}
f_\alpha (E)=\{\exp [(E-\mu _\alpha )/kT]+1\}^{-1}.  \label{12}
\end{equation}

The noise spectra of both charge-current and spin-current can be obtained
from the correlation $S_{\alpha \beta }^{\sigma \sigma ^{\prime }}$ between
spin-dependent particle currents in lead $\alpha $ and $\beta $: 
\begin{equation}
S_{\alpha \beta }^{\sigma \sigma ^{\prime }}=\left\langle \left[ \overset{%
\wedge }{J}_{\alpha \sigma }\left( t_1\right) -\left\langle \overset{\wedge }%
{J}_{\alpha \sigma }\left( t_1\right) \right\rangle \right] \left[ \overset{%
\wedge }{J}_{\beta \sigma ^{\prime }}\left( t_2\right) -\left\langle \overset%
{\wedge }{J}_{\beta \sigma ^{\prime }}\left( t_2\right) \right\rangle \right]
\right\rangle  \label{13}
\end{equation}
Here $\left\langle \cdots \right\rangle $ denotes both statistical average
and quantum average on the nonequilibrium state.

Briefly, we substitute Eq.(7) and Eq.(9) into Eq.(13) and apply the analytic
continuation so that the exact expression of the zero-frequency,
spin-dependent correlation can be obtained. It has been shown in Ref.\cite%
{23} that both the cross- and the auto-correlation are needed to
characterize the shot noise of spin-current for the two-lead system because
a spin-current is not conserved due to the spin flip induced by the rotating
magnetic field. The cross correlation shot noise of spin-current is 
\begin{eqnarray}
S_{LR} &\equiv &S_{spin,1}=\left\langle (\vartriangle J_{L\uparrow
}-\vartriangle J_{L\downarrow })(\vartriangle J_{R\uparrow }-\vartriangle
J_{R\downarrow })\right\rangle  \nonumber \\
&=&\int \frac{dE}{8\pi }f_{\downarrow }(1-f_{\uparrow })\Gamma _L\Gamma
_R[\left| G_{\uparrow \downarrow }^r\right| ^2+\left| G_{\downarrow \uparrow
}^r\right| ^2-2\Gamma ^2(\left| G_{\uparrow \downarrow }^r\right| ^4+\left|
G_{\downarrow \uparrow }^r\right| ^4)]  \label{14}
\end{eqnarray}
and the auto-correlation shot noise is defined by

\begin{eqnarray}
S_{LL} &\equiv &S_{spin,2}=\left\langle (\vartriangle J_{L\uparrow
}-\vartriangle J_{L\downarrow })(\vartriangle J_{L\uparrow }-\vartriangle
J_{L\downarrow })\right\rangle  \nonumber \\
&=&\int \frac{dE}{8\pi }f_{\downarrow }(1-f_{\uparrow })\{2[-\Gamma
_L^2\Gamma ^2(\left| G_{\uparrow \downarrow }^r\right| ^4+\left|
G_{\downarrow \uparrow }^r\right| ^4)+\Gamma _L\Gamma (\left| G_{\uparrow
\downarrow }^r\right| ^2+\left| G_{\downarrow \uparrow }^r\right| ^2)] 
\nonumber \\
&&-\Gamma _L\Gamma _R(\left| G_{\uparrow \downarrow }^r\right| ^2+\left|
G_{\downarrow \uparrow }^r\right| ^2)\},  \label{15}
\end{eqnarray}
where $f_{\uparrow }=f_{\uparrow }(E)$ and $f_{\downarrow }=f_{\downarrow
}(E-w).$

Once the retarded Green's functions $G_{\sigma \sigma ^{\prime
}}^{r(a)}(E_1,E_2)$ are known, the spin current and shot noise can be
calculated using Eqs. (10) and (13-15). In the following we calculate $%
G_{\sigma \sigma ^{\prime }}^{r(a)}(E_1,E_2)$ with the standard Dyson
equation approach. To this end we regard the term, which explicitly depends
on time $\mathbf{t}$, in the Hamiltonian Eq.(1) as the interacting part $H_I$
such that $H_0\equiv H-H_I$ . Denoting Green's functions for the Hamiltonian 
$H_0$ as $G_{\sigma \sigma ^{^{\prime }}}^{0r}(\varepsilon )$ the full
Green's functions for Hamiltonian Eq.(1) are then calculated from the Dyson
equation

\begin{eqnarray}
\mathbf{G}^r(E_1,E_2) &=&2\pi \mathbf{G}^{0r}(E_1)\delta (E_1-E_2)  \nonumber
\\
&&+\int \frac{dE}{2\pi }\mathbf{G}^r(E_1,E+E_2)\mathbf{H}^{\prime }(E)%
\mathbf{G}^{0r}(E_2),  \label{16}
\end{eqnarray}
where the boldface notation indicates that the electron Green's function in
the QD and the interacting Hamiltonian $\mathbf{H}^{\prime }$ are the $%
2\times 2$ matrices in the spin space , where the element $H_{\sigma
_1\sigma _2}^{\prime }(E)$ is the Fourier transformation of $H_{\sigma
_1\sigma _2}^{\prime }(t)$ which is seen to be

\begin{equation}
H_{\sigma _1\sigma _2}^{\prime }(t)=r[e^{-iwt}d_{\uparrow }^{+}d_{\downarrow
}+e^{iwt}d_{\downarrow }^{+}d_{\uparrow }].  \label{17}
\end{equation}
The full retarded Green's functions of Hamiltonian (1) are then obtained
from Eq. (16) , after tedious but straightforward algebra, explicitly as\cite%
{24}

\begin{equation}
G_{\sigma \sigma }^r(E_1,E_2)=\frac{2\pi \delta (E_1-E_2)G_{\sigma \sigma
}^{0r}(E_1)}{1-r^2g(E_1)},  \label{18}
\end{equation}

\begin{equation}
G_{\sigma \overline{\sigma }}^r(E_1,E_2)=2\pi \delta (E_1+\overline{\sigma }%
w-E_2)\frac{rg(E_1)}{1-r^2g(E_1)}  \label{19}
\end{equation}
where $g(E_1)\equiv G_{\sigma \sigma }^{0r}(E_1)G_{\sigma \overline{\sigma }%
}^{0r}(E_1+\overline{\sigma }w),\overline{\sigma }=-\sigma $. Using these
relations, it is straightforward to derive the expression of the
spin-current and shot noise from Eq.(10) and Eq.(13-15).

Since we are interested in the case with strong strength of E-PH
interaction, the Green's function $G_{\sigma \sigma ^{^{\prime }}}^{0r}(E)$
can be calculated by performing a standard canonical transformation, $%
\overset{\thicksim }{H}=e^sHe^{-s}$ with $S=(\frac \lambda {w_0})\underset{%
\sigma }{\sum }d_\sigma ^{+}d_\sigma (a^{+}-a)$\cite{39}. And then all
transformed operators are seen to be

\begin{eqnarray}
\overset{\thicksim }{d}_\sigma &=&d_\sigma X  \nonumber \\
\overset{\thicksim }{d}_\sigma ^{+} &=&d_\sigma ^{+}X^{+}  \label{20} \\
\overset{\thicksim }{a} &=&a-\frac \lambda {w_0}\underset{\sigma }{\sum }%
d_\sigma ^{+}d_\sigma  \nonumber \\
\overset{\thicksim }{a}^{+} &=&a^{+}-\frac \lambda {w_0}\underset{\sigma }{%
\sum }d_\sigma ^{+}d_\sigma  \label{21}
\end{eqnarray}
with

\begin{equation}
X=\exp [-\frac \lambda {w_0}(a^{+}-a)].  \label{22}
\end{equation}
The electron number operator in the QD is invariant under the transformation,

\begin{equation}
\overset{\thicksim }{d}_\sigma ^{+}\overset{\thicksim }{d}_\sigma =d_\sigma
^{+}d_\sigma .  \label{23}
\end{equation}
Then the transformed Hamiltonian can be written as

\begin{equation}
\overset{\thicksim }{H}=\overset{\thicksim }{H}_L+\overset{\thicksim }{H}_R+%
\overset{\thicksim }{H}_X+\overset{\thicksim }{H}_D+\overset{\thicksim }{%
H^{\prime }}(t)+\overset{\thicksim }{H}_T  \label{24}
\end{equation}
where

\[
\overset{\thicksim }{H}_L+\overset{\thicksim }{H}_R=\underset{k\sigma
,\alpha =L,R}{\sum }\varepsilon _kC_{k\alpha \sigma }^{+}C_{k\alpha \sigma } 
\]
\[
\overset{\thicksim }{H}_X=w_0(a^{+}-\frac \lambda {w_0}\underset{\sigma }{%
\sum }d_\sigma ^{+}d_\sigma )(a-\frac \lambda {w_0}\underset{\sigma }{\sum }%
d_\sigma ^{+}d_\sigma ) 
\]
\[
\overset{\thicksim }{H}_D=\underset{\sigma }{\sum }[(\varepsilon +\sigma
B_0\cos \theta )+\lambda (a+a^{+}-2\frac \lambda {w_0}\underset{\sigma
^{\prime }}{\sum }d_{\sigma ^{\prime }}^{+}d_{\sigma ^{\prime }})]d_\sigma
^{+}d_\sigma 
\]
\[
\overset{\thicksim }{H^{\prime }}(t)=r[\exp (-iwt)d_{\uparrow
}^{+}d_{\downarrow }+\exp (iwt)d_{\downarrow }^{+}d_{\uparrow }] 
\]
\begin{equation}
\overset{\thicksim }{H}_T=\underset{k,\sigma ,\alpha =L,R}{\sum }[T_{k\alpha
}C_{k\alpha \sigma }^{+}d_\sigma X+c.c].  \label{25}
\end{equation}
The hopping terms between molecule and leads [ Eq. (25)] is modified by a
factor $X,$ which describes the fact that the electron hopping is
accompanied by a phonon cloud. Here to avoid unnecessary complication, we
consider the leads which are unaffected by the phonons. This means that we
ignore a factor which results from the average of the $X$ operator and does
not lead to qualitative changes of the tunneling current. The justification
for this is given in Ref. 16 and Ref. 39. Consequently we have 
\[
\overset{\thicksim }{H}_T\approx H_T 
\]
and 
\begin{equation}
\overset{\thicksim }{H}=\overset{\thicksim }{H_{el}}+\overset{\thicksim }{H}%
_{ph}  \label{26}
\end{equation}
where 
\begin{eqnarray*}
\overset{\thicksim }{H_{el}} &=&\underset{k\sigma ,\alpha =L,R}{\sum }%
\varepsilon _kC_{k\alpha \sigma }^{+}C_{k\alpha \sigma }+\underset{\sigma }{%
\sum }(\varepsilon +\sigma B_0\cos \theta -\vartriangle )d_\sigma
^{+}d_\sigma \\
&&+\underset{k,\sigma ,\alpha =L,R}{\sum }[T_{k\alpha }C_{k\alpha \sigma
}^{+}d_\sigma +c.c]+r[\exp (-iwt)d_{\uparrow }^{+}d_{\downarrow }+\exp
(iwt)d_{\downarrow }^{+}d_{\uparrow }]
\end{eqnarray*}
\[
\overset{\thicksim }{H}_{ph}=w_0a^{+}a 
\]
with $\vartriangle =\frac{\lambda ^2}{w_0}$. Due to the E-PH interaction,
the single energy level of the molecule is renormalized to $\varepsilon
^{\prime }=\varepsilon -\vartriangle $ and then the Green's function ,$%
G_{\sigma \sigma ^{^{\prime }}}^{0r}(t)$ , can be decoupled as 
\begin{equation}
G_{\sigma \sigma ^{\prime }}^{0r}(t)=-i\theta (t)\left( \left\langle \overset%
{\thicksim }{d}_\sigma (t)\overset{\thicksim }{d}_{\sigma ^{\prime
}}^{+}(0)\right\rangle _{el}\left\langle X(t)X^{+}(0)\right\rangle
_{ph}-\left\langle \overset{\thicksim }{d}_{\sigma ^{\prime }}^{+}(0)\overset%
{\thicksim }{d}_\sigma (t)\right\rangle _{el}\left\langle
X^{+}(0)X(t)\right\rangle _{ph}\right)  \label{27}
\end{equation}
where 
\[
\overset{\thicksim }{d}_\sigma (t)=e^{i\overset{\thicksim }{H}%
_{el}t}d_\sigma e^{-i\overset{\thicksim }{H}_{el}t}, 
\]
\[
X(t)=e^{i\overset{\thicksim }{H}_{ph}t}Xe^{-i\overset{\thicksim }{H}_{ph}t}. 
\]
The renormalization factor due to the E-PH interaction is evaluated as \cite%
{39}: 
\[
\left\langle X(t)X^{+}(0)\right\rangle _{ph}=e^{-\Phi (t)} 
\]
\[
\left\langle X^{+}(0)X(t)\right\rangle _{ph}=e^{-\Phi (-t)} 
\]
where 
\begin{equation}
\Phi (t)=g[N_{ph}(1-e^{iw_0t})+(N_{ph}+1)(1-e^{-iw_0t})]  \label{28}
\end{equation}
with parameters $N_{ph}=\frac 1{\exp (\beta w_0)-1}$ and $g=(\frac \lambda
{w_0})^2$.

The Fourier transform of the Green's function $G_{\sigma \sigma ^{^{\prime
}}}^{0r}(E)$ is given by : 
\begin{eqnarray}
G_{\sigma \sigma }^{0r}(E) &=&\exp [-g(2N_{ph}+1)]\overset{\infty }{\underset%
{l=-\infty }{\sum }}I_l[2g\sqrt{N_{ph}(N_{ph}+1)}]\exp ^{(lw_0/2)\beta }\exp
^{-iw_0lt}  \nonumber \\
&&*[(1-\left\langle n_{d,\sigma }\right\rangle )\overset{\thicksim }{G}%
_{\sigma \sigma }^{0r}(E-lw_0)+\left\langle n_{d,\sigma }\right\rangle 
\overset{\thicksim }{G}_{\sigma \sigma }^{0r}(E+lw_0)]  \label{29}
\end{eqnarray}
where $\overset{\thicksim }{G}_{\sigma \sigma }^{0r}(E)$ is the retarded
Green function corresponding to the time-independent part of the new
Hamiltonian $\overset{\thicksim }{H_{el}}$ , the index $l$\textit{\ }%
indicates the number of phonons involved, and $\left\langle n_{d,\sigma
}\right\rangle $ is the time-averaged electron occupation number in the
molecule. Here we consider the case of an ''empty'' QD i.e. $\left\langle
n_{d,\sigma }\right\rangle $ =0. With a little algebra we find 
\begin{equation}
\overset{\thicksim }{G}_{\sigma \sigma ^{\prime }}^{0r}(E)=\frac{\delta
_{\sigma \sigma ^{\prime }}}{E-(\varepsilon +\sigma B_0\cos \theta
-\vartriangle )-\Sigma _{\sigma \sigma ^{\prime }}^r}  \label{30}
\end{equation}
where the retarded self-energy due to the tunneling into the electrical
leads are given by 
\begin{equation}
\Sigma _{\sigma \sigma ^{\prime }}^r(E)=\underset{k,\alpha \in L,R}{\sum }%
\frac{\left| T_{k\alpha }\right| ^2\delta _{\sigma \sigma ^{\prime }}}{%
E-\varepsilon _k+i0^{+}}=\Lambda (E)-\frac i2\Gamma (E).  \label{31}
\end{equation}
In the wide-band limit, the level shift $\Lambda (E)$ can be neglected and
the linewidths are energy independent constants. Thus the retarded
self-energy can be expressed as 
\begin{equation}
\Sigma _{\sigma \sigma ^{\prime }}^r(E)=-\frac i2\Gamma .  \label{32}
\end{equation}
The Fourier transform of the full Green's function given by Eq. (27) can be
obtained as 
\begin{eqnarray}
G_{\sigma \sigma ^{\prime }}^{0r}(E) &=&\exp [-g(2N_{ph}+1)]  \label{33} \\
&&*\overset{\infty }{\underset{l=-\infty }{\sum }}I_l[2g\sqrt{%
N_{ph}(N_{ph}+1)}]\frac{e^{w_0l\beta /2}}{E-(\varepsilon +\sigma B_0\cos
\theta -\vartriangle )-lw_0+\frac i2\Gamma }  \nonumber
\end{eqnarray}

\section{NUMERICAL RESULTS AND DISCUSSIONS}

We now present the numerical results of the spin current and zero-frequency
shot noise. For simplicity, we consider the symmetric tunnel-coupling
between the molecule and the two leads i.e. $\Gamma _L=\Gamma _R=\frac
\Gamma 2$ and further assume that the energy level of the molecule which is
controlled by the gate voltage $V_g$ such that $\varepsilon
(V_g)=\varepsilon _0+eV_g$, where $\varepsilon _0$ denotes\textbf{\ }%
single-electron energy in the molecule in the absence of the gate voltage\ .
The phonon energy\ is chosen as the energy unit throughout the rest of the
paper. We also set $\hbar =e=1$. In fig.(2) we plot the spin current $I_s$
versus the parameter $r$ (Fig.(2a)), the gate voltage $V_g$ (Fig.(2b)) and
frequency $w$ of the rotating magnetic field respectively(Fig.(2c)) at zero
temperature $T=0.$ The parameter values used in Fig.(2a) are such that $%
\Gamma =0.04,V_g=0,g=0.5^2$ and $\theta =88^{\circ }$. For comparison, we
also plot the spin current in the absence of E-PH interaction(dashed line).
It is clearly shown in Fig.(2a) that the E-PH interaction results in the
shift of resonance peaks. Fig.(2b) shows the spin-current $I_s$ as a
function of the gate voltage $V_g$ for different $r$ with $\Gamma
=0.04,g=0.6^2,w=0.1$and $\theta =88^{\circ }$ \textbf{. }In the presence of
E-PH coupling the overall spectrum is shifted by a quantity $\vartriangle =%
\frac{\lambda ^2}{w_0}$ toward the negative gate voltage region$.$ In
addition to the main peak which is related to the molecule energy level,
small satellite resonant peaks appear at the positive energy side. At zero
temperature only phonon-emission processes are allowed since before
tunneling the phonon state is vacuum, which explains why the satellite peaks
are located at the positive energy region. Moreover, the height of the
satellite peaks is much smaller than the main resonant peak because of the
suppression by the E-PH coupling. Fig.(2c) display the spin current $I_s$
versus frequency $w$ of the rotating magnetic field for various coupling
constants $g=0,0.3^2,0.5^2,0.7^2$ (correspondingly the solid, dotted ,
dashed and the dash-dotted curves)$,$with $\Gamma =0.04,$ $V_g=0,$ $%
r=0.08,\theta =88^{\circ }.$ In the absence of the phonon mode, there exists
only one resonant peak. The E-PH coupling typically leads to the new
satellite resonant peaks. It is also seen that the heights of satellite
peaks increase with the coupling constant $g$. The positions of the side
peaks are located in $w=nw_0$ $\left( n=1,2,3,4\cdots \right) .$ Finally,
the plot of spin current $I_s$ as a function of E-PH coupling constant $g$ (
Fig.(2d)) shows a double-maximum. Fig.(3) shows the dependence of
cross-correlation shot noise on the parameter $r$ (Fig.(3a)), the gate
voltage $V_g$ (Fig.(3b)) and frequency $w$ of the rotating magnetic
field(Fig.(3c)) respectively with the same parameter values as in Fig.(2).
We see from Fig.(3) that the cross-correlation shot noise of the
spin-current displays very different behavior from the spin-current itself,
therefore the measurements of the shot noise spectrum can provide more
information of the transport properties in mesoscopic systems. From Fig.(3a)
it is observed that in the absence of E-PH interaction the cross-correlation
shot noise can be either positive or negative as $r$ is changed due to the
competition\cite{27} between the cross correlations of electrons with
parallel and antiparallel spins. The E-PH interaction has considerable
influence on the\textbf{\ }cross-correlation shot noise that the negative
shot noise is substantially suppressed and, even vanishes. Fig.(3b) shows
that with $r$ increasing, the cross-correlation shot-noise spectra exhibits
two extra peaks located symmetrically around the position of the main peak,
and in the case of stronger magnetic field, the cross-correlation \ shot
noise turns to negative for some gate voltages while it remains positive for
the other gate voltages. From Fig.(3c) the oscillating behavior of the shot
noise between positive and negative values is observed and is due to the
photon assisted process. We finally plot the auto-correlation shot noise
versus the gate voltage $V_g$ ( Fig.(4a) ) and frequency $w$ ( Fig.(4b)) in
Fig.4. The auto correlation shot noise spectra are also shifted and
satellite peak appears compared with the absence of the E-PH interaction.
Moreover, the auto-correlation of the spin-current is surely positive
definite which is different from the situation for cross-correlation. We see
that the shot noises possess reach informations of the phonon effect on the
quantum transport through molecular devices.

\section{CONCLUSIONS}

In summary we have shown for the first time that the shot noise of spin
current in a single-molecule quantum dot can be significantly affected by
the phonon mode and can provide a beneficial information to improve the
understanding of transport properties through the molecular QDs. It is shown
that in addition to the shift of the resonant-peak position associated with
the level of the dot, satellite peaks emerge at integer number of the phonon
frequency. The E-PH coupling even can reverse the sign of the zero-frequency
cross shot noise.

\section{Acknowledgment}

This work was supported by National Natural Science Foundation of China
under Grant Nos.10475053.

Figure Captions:

FIG.1. Schematic diagram of a molecule quantum-dot system.

FIG. (2a)The spin current $I_s$ versus the parameter $r$ ( $\Gamma
=0.04,V_g=0,g=0.5^2$ , $\theta =88^{\circ }$ with frequencies of magnetic
field $w=0.05,0.25,0.7$ and $1.5$ ) for the two cases with (solid line) and
without the phonon mode ( dashed line) as comparison\textbf{.}

FIG. (2b) The spin current $I_s$ as a function of the gate voltage $V_g$ ( $%
\Gamma =0.04,g=0.6^2,w=0.1$ and $\theta =88^{\circ }$ with the parameters $%
r=0.02,0.04,0.06$ and $0.08$) for the two cases with (solid line) and
without the phonon mode ( dashed line) as comparison\textbf{.}

FIG.(2c) The spin current $\frac{I_s}w$ versus frequency $w$ of the rotating
magnetic field with $\Gamma =0.04,$ $V_g=0,$ $r=0.08,\theta =88^{\circ }$
for various coupling constants $g=0$ (solid curve) $g=0.3^2$(dotted curve) $%
g=0.5^2$(dashed curve) $g=0.7^2$ (dash-dotted curve).

FIG. (2d) The spin current $I_s$ as a function of coupling constant $g$ .

FIG. (3a) The cross- correlation shot noise as a function of the parameter $%
r $ (solid line: with phonon mode; dashed line: without phonon mode).

FIG. (3b) The cross-correlation shot noise as a function of the gate voltage 
$V_g$ (solid line: with phonon mode; dashed line: without phonon mode).

FIG. (3c) The cross-correlation shot noise $\frac{S_{LR}}w$ versus frequency 
$w$ for different coupling constants: $g=0$ (solid curve) $g=0.3^2$(dotted
curve) $g=0.5^2$(dashed curve) $g=0.7^2$ (dash-dotted curve) with $\Gamma
=0.04,$ $V_g=0,$ $r=0.08,\theta =88^{\circ }.$

FIG. (4a) The auto-correlation shot noise versus the gate voltage $V_g$ with 
$r=0.02,0.06,0.1$and $0.2$ (solid line: with phonon mode; dashed line:
without phonon mode).

FIG.(4b) The auto-correlation shot noise versus frequency $w$ with the same
parameters as in Fig.(2c) for various coupling constants $g=0$ (solid curve) 
$g=0.3^2$(dotted curve) $g=0.5^2$(dashed curve) $g=0.7^2$ (dash-dotted curve)%
\textbf{.}


\begin{thebibliography}{99}
\bibitem{1} A. Aviram and M. Ratner,eds.,\textit{Molecular Electronics:
Science and Technology} (Annals of the New York Academy of Science,New
York,1998).

\bibitem{2} V. Langlais \textit{et al}., Phys. Rev. Lett. \textbf{83}, 2809
(1999).

\bibitem{3} B. C. Stipe, M. A. Rezaei, W. Ho, S. Gao, M. Persson, and B. I.
Lundqvist, Phys. Rev. Lett. \textbf{78,} 4410 (1997)

\bibitem{4} D. Porath and O. Milo, J. Appl. Phys. \textbf{81}, 2241 (1997)

\bibitem{5} E. W. Wong \textit{et al}., J. Am. Chem. Soc. \textbf{122}, 5831
(2000)

\bibitem{6} M. A. Reed, C. Zhou, C. J. Muller, T. P. Burgin, and J. M. Tour,
Science \textbf{278}, 252 (1997)J. Chen, M. Reed, A. Rawlett, and J. Tour,
Science\textbf{\ 286}, 1550 (1999).

\bibitem{7} M. Ventra, S. G. Kim. S. Pantelides, and N. Lang, Phys. Rev.
Lett. \textbf{86}, 288 (2001).

\bibitem{8} H. Park, A. Lim, e. Anderson, A. Allvisatos, and P. McEuen,
Nature \textbf{407}, 57 (2000); J, Park \textit{et al., }Nature\textbf{\ 417}%
, 722 (2002); J. Recichert \textit{et al., }Phys. Rev. Lett. \textbf{88},
176804 (2002)

\bibitem{9} L. H. Yu and D. Natelson, Nano Letters, \textbf{4}, 79 (2004).

\bibitem{10} W. Liang, M. P. Shores, M. Brockrath, J. R. Long, and H. Park,
Nature \textbf{417}, 725 (2002).

\bibitem{11} N. B. Zhitenev, H. Meng, and Z. Bao, Phys. Rev. Lett. \textbf{88%
}, 226801 (2002).

\bibitem{12} X. H. Qiu, G. V. Nazin, and W. Ho, Phys. Rev. Lett. \textbf{92}%
,206102 (2004)

\bibitem{13} A. R. Champagne, A. N. Pasupathy, and D. C. Ralph,
cond-mat/0409134 (2004)

\bibitem{14} E. Emberly and G. Kirczenow, Phys. Rev. B\textbf{\ 61}, 5740
(2000)

\bibitem{15} D. Bose and H. Schoeller, Europhys. Lett. \textbf{54},668
(2001); A. S. Alexandrov and A. M. Bratkovsky, Phys. Rev. B \textbf{67},
235312 (2003); K. D. McCarthy, N. Prokof'evev, and M. T. Tuominen, Phys.
Rev. B \textbf{67} 245415 (2003).

\bibitem{16} U. Lundin and R. H. McKenzie, Phys. Rev.B \textbf{66},75303
(2002).

\bibitem{17} J. X. Zhu and A. V. Balatsky, Phys. Rev. B \textbf{67}, 165326
(2003).

\bibitem{18} K. Flensberg, Phys. Rev. B \textbf{68}, 205323 (2003).

\bibitem{19} V. Aji, J. E. Moore, and C. M. Varma, cond-mat/0302222 (2003)

\bibitem{20} J. Koch, F. von Pooen, Y. Oreg, and E. Sela, cond-mat/0405433
(2004)

\bibitem{21} S. Braig and K. flenberg, Phys. Rev. B\textbf{\ 68}, 205324
(2003).

\bibitem{22} A. Mitra, I. Aleiner, and A. J. Millis, Phys. Rev. B \textbf{69}%
,245302 (2004).

\bibitem{23} B.G. Wang, J. Wang, and H. Guo, cond-mat/0305066 (2003)

\bibitem{24} E.G. Emberly and G. Kirezenow, Phys. Rev. B \textbf{62},10451
(2000)

\bibitem{25} N. S. Wingreen, K. W. Jacobsen, and J.W. Wilkins, Phys. Rev. B 
\textbf{40}, 11834 (1989)

\bibitem{26} H. A. Engel and D. Loss, Phys. Rev. B \textbf{65}, 195321
(2002); D. Mozyrsky et al., Phys. Rev. B \textbf{66}, 161313 (2002)

\bibitem{27} J. X. zhu and A. V. Balatsky, Phys. Rev. Lett. \textbf{89},
286802 (2002)

\bibitem{28} I. Martin et al.,Phys. Rev. Lett. \textbf{90}, 18301 (2003)

\bibitem{29} S. A. Wolf et al., Science \textbf{294}, 1488 (2001); G.A.
Prinz, Science \textbf{282}, 1660 (1998)

\bibitem{30} M. J. Stevens et al.,Phys. Rev. Lett \textbf{90},136603 (2003);
J. H$\ddot{u}$bner et al., Phys. Rev. Lett, \textbf{90}, 216601, (2003);
Susan K. Watson \textit{et al}., cond-mat/0302492 (2003)

\bibitem{31} A. Brataas \textit{et. al.}, Phys. Rev. B\textbf{\ 66}, 060404
(2002)

\bibitem{32} B. G. Wang, J. Wang, and H. Guo, Phys.Rev. B\textbf{\ 67},
92408 (2003)

\bibitem{33} P. Sharma and C. Chamon, Phys. Rev. Lett. \textbf{87}, 096401
(2001); E.R. Mucciolo\textit{\ et al.}, ibid. \textbf{89}, 146802 (2002);
Q.-f. Sun, H.Guo, and J. Wang, cond-mat/0212293 (2002)

\bibitem{34} B.G. Wang, J. Wang, and H. Guo. cond-mat/0305066 (2003)

\bibitem{35} J. R. Pierce, Bell Syst. Tech. J. \textbf{27},158 (1948)

\bibitem{36} The z-component of field B(t) splits the level $\epsilon $ into
two levels with $\epsilon _{\downarrow }\prec \epsilon _{\uparrow }$. The
chemical potentials $\mu $ of the leads is adjusted by the gate voltage so
that $\epsilon _{\downarrow }\prec \mu \prec \epsilon _{\uparrow }$ and no
bias voltage is applied to the two leads. A spin-down electron can tunnels
into state $\epsilon _{\downarrow \text{ }}$from the left lead and absorbs a
photon to flip its spin and occupy the state $\epsilon _{\uparrow }.$
Because $\epsilon _{\uparrow }\succ \mu ,$ it is more easy for the spin-up
electron to tunnel out to the leads. The same process occurs for spin-down
electrons coming from the right lead. Therefore spin-down electrons flow
toward to the quantum dot while spin-up electrons flows away from the dot
giving rise to a zero charge-current and a net spin-current.

\bibitem{37} A. P. Jauho et al, Phys. Rev. B \textbf{50}, 5528 (1994)

\bibitem{38} D.C. Langreth, in \textit{Liner and Nonlinear Electron
Transport in Solids}, Vol. 17 of \textit{Nato Advanced Study Institute,
Series B: Physics}, edited by J. T. Devreese and V. E. Van Doren

\bibitem{39} See, e.g., G. D. Mahan, \textit{Many-particle Physics}, 2nd ed.
(Plenum Press, New York, 1990), pp. 285-324

\bibitem{40} A. Hewson and D. Newns, J. Phys. C \textbf{13}, 4477 (1980)

\bibitem{41} Ya. M. Blanter and M. Buttiker, Phys. Rep. \textbf{336}, 1
(2000)

\bibitem{42} M. Buttiker, Phys. Rev. B \textbf{46},12485 (1992)

\bibitem{43} E. M. Prucell, Nature, \textbf{178}, 1449 (1956)

\bibitem{44} L. V. Keldysh, Zh. Eksp.Teor. Fiz.. \textbf{47}, 1515 (1965);C.
Caroli \textit{et al., }J. Phys. C\textbf{\ 4}, 916 (1971)
\end{thebibliography}
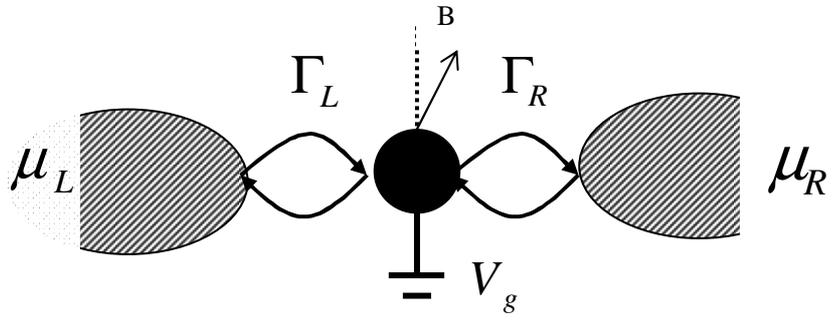
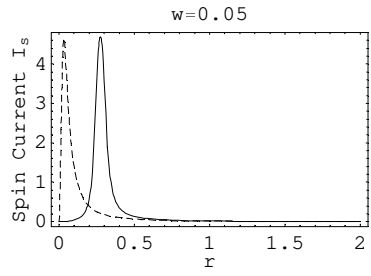
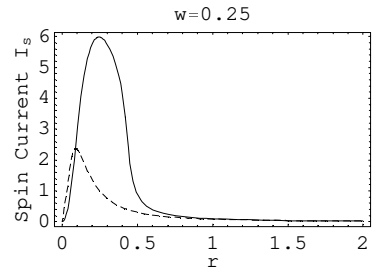
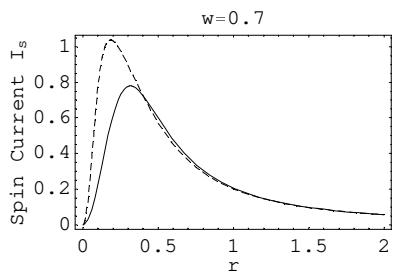
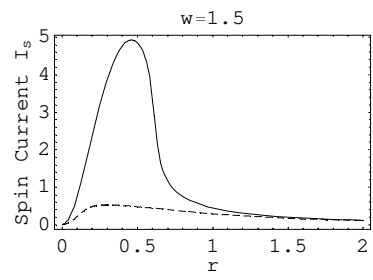
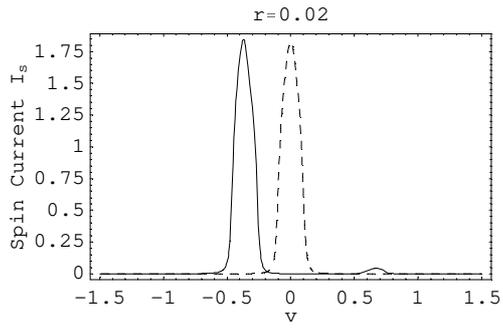
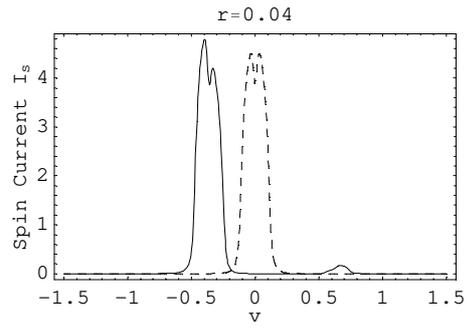
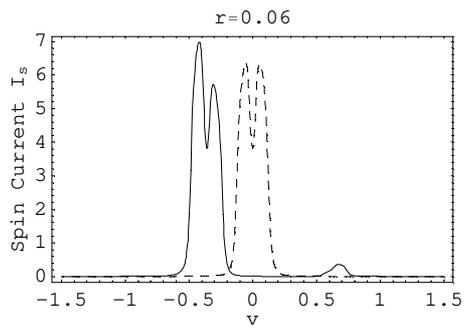
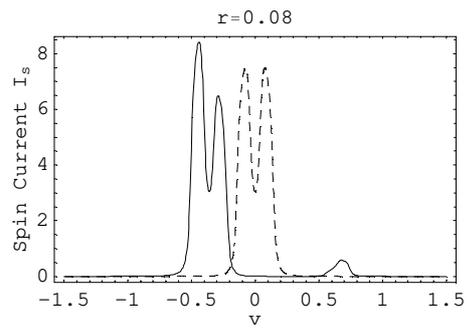
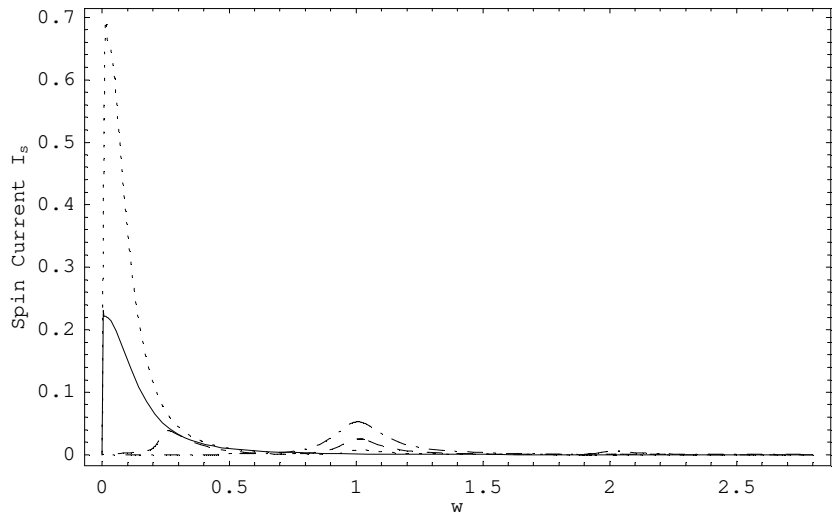
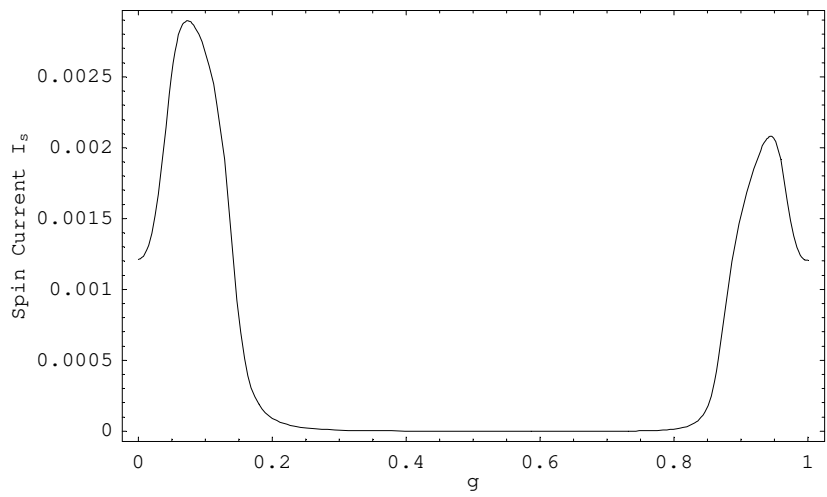
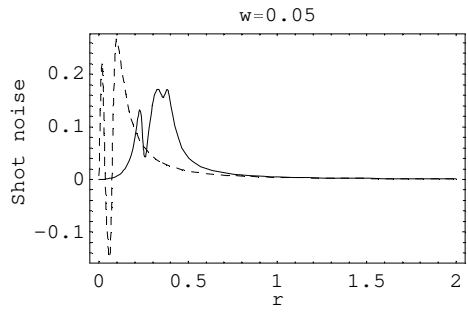
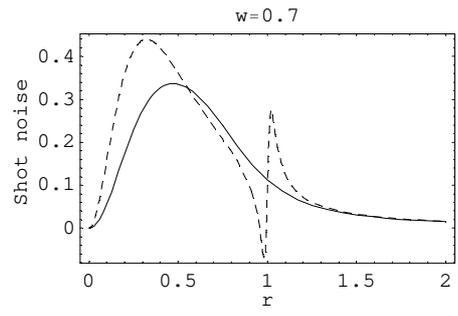
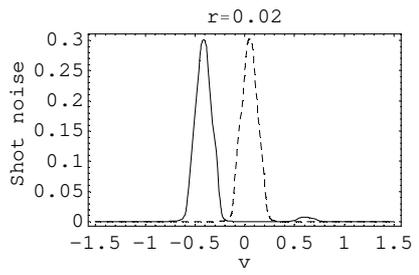
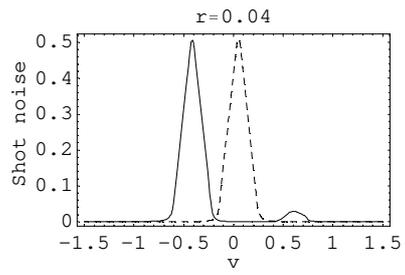
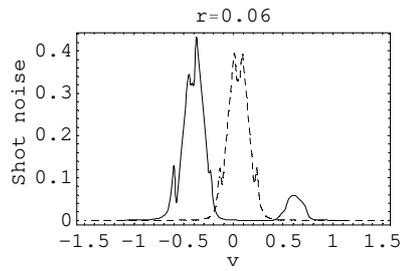
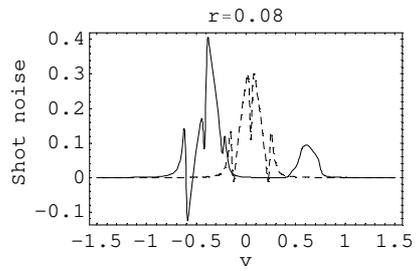
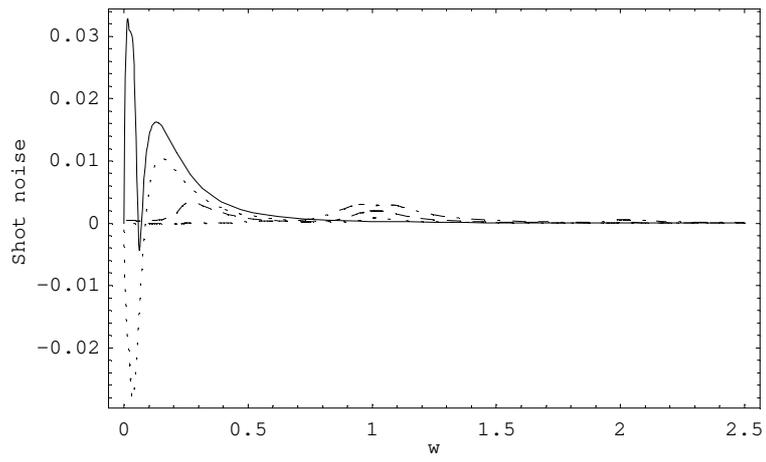
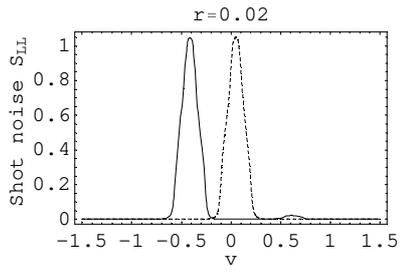
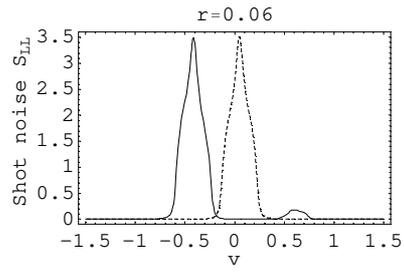
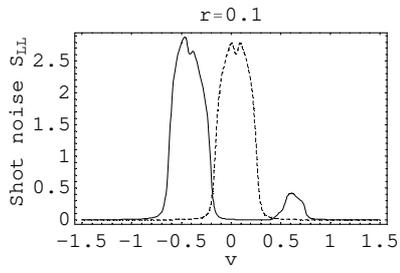
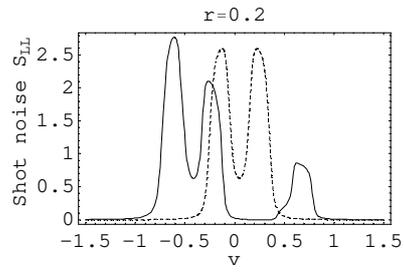
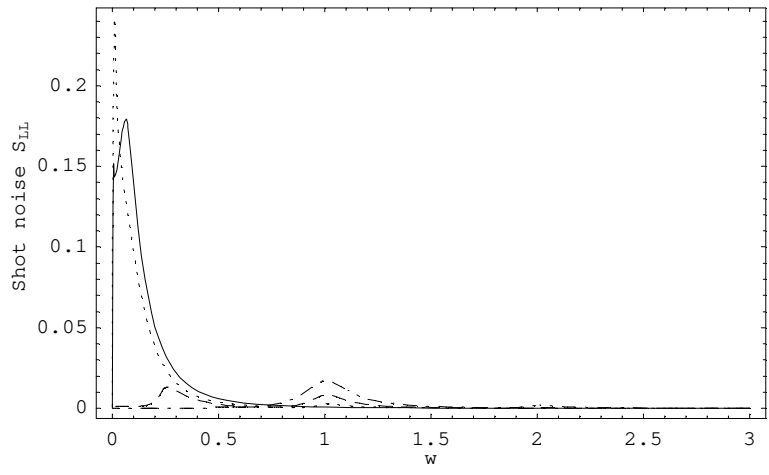
\end{document}